# Hot bubble and slow wind dynamics in PNe, Radiation-gasdynamics of PNe V

Garrelt Mellema[1] and Adam Frank[2]

[1] *Astrophysics Group, Department of Mathematics, UMIST, P.O. Box 88, Manchester M60 1QD, UK*
[2] *Department of Astronomy, University of Minnesota, 116 Church St. S.E., Minneapolis MN 55455, USA*

**ABSTRACT**
This paper looks into various aspects brought to light by numerical work on the generalized interacting winds model for planetary nebulae. First, a detailed comparison between radiative and non-radiative models is made, showing that one's naive expectations of the effects of radiative heating and cooling are not always true. Secondly, we consider the evolution of the slow wind after it has gotten ionized. It is found that the initial aspherical density distribution is smoothed out after ionization, thus requiring the aspherical nebula to be formed within the first few thousand years of PN evolution. Thirdly, the nature of the flow in the hot bubble is investigated. Both one and two-dimensional numerical models show strong signs of instabilities or turbulent flow in the hot bubble. Although observationally hard to prove or disprove this turbulent structure is critically examined. It is found that although the turbulence is not a numerical artifact, the full three-dimensional picture will most definitely differ from what is found in two dimensions. The implications for the interpretation of the models are discussed. Finally, the issue of the soft X-ray emission from PNe is considered. It is found that soft X-rays originate mainly from the thin interface between the hot bubble and the actual nebula.

## 1 INTRODUCTION

The generalized interacting winds model (GISW) has established itself as the standard model for explaining the shapes and shaping of planetary nebulae (PNe). Extending the original interacting winds model for PNe of Kwok et al. (1978) by introducing an axi-symmetric slow wind has proved to be quite successful, and has revealed many interesting aspects of PN gasdynamics (Kahn & West 1985; Balick 1987; Soker & Livio 1989; Mellema et al. 1991; Icke et al. 1992)

The current series of papers has concentrated on studying the GISW model under radiative circumstances using the method described in Frank & Mellema (1994a) (Paper I). Frank (1994) and Mellema (1994a) (Papers II and III) looked at one-dimensional spherical models, mainly concentrating on effects introduced by the evolution of the stellar radiation field and/or the accelerating fast wind. Paper IV (Frank & Mellema 1994b) provided a collection of aspherical models presented in the form of synthesized images and long slit spectra in prominent emission lines. This collection gives a good impression of the shapes and kinematics that can be produced by the GISW model. The comparisons to observed images and spectra showed that the results of the GISW simulations do indeed describe the situation in aspherical PNe very well.

In this paper we address some issues that are relevant for the GISW models, and were not treated in Paper IV. After a short summary of the character of our numerical models and initial/boundary conditions in Sect. 2, we begin with a detailed comparison between radiative and non-radiative models (Sect. 3). Section 4 addresses the effects photo-ionization has on the slow wind. After that we look into the turbulent state of the hot bubble that was found in the simulations, especially addressing the numerical versus physical character of this phenomenon (Sect. 5). Section 6 deals with the soft X-ray emission that can be predicted from our models. The last section gives a summary of the results in this paper.

## 2 NUMERICAL APPROACH

The code used to do the calculations in this article is the combination of the two-dimensional numerical gasdynamics solver described in Mellema et al. (1991) (a Roe solver, see also Eulderink 1993), and the radiation method described in Paper I. This method adds time-dependent ionization/recombination and heating/cooling to the gasdynamic solver. It uses the local radiation field, found from solving the radiation transport problem, to calculate the ionization fractions and the heating and cooling terms. Ionization is caused by photons and collisions. Heating is from photo-ionization of H and He. Cooling in the 'nebular' regime ($T < 10^5$ K) is caused by recombinations of H and He, a range of collisionaly excited lines of $C^{+3}$, $N^{+1}$, $O^+$, $O^{+2}$, H, and He. In the 'coronal' regime ($T > 10^5$ K) a cooling curve from the literature is used.

The merits of the gasdynamic solver used were discussed in Mellema et al. (1991). Because it uses characteristics, it is particularly suited for a problem like the interacting winds model. This problem is characterized by large jumps in density, velocity and pressure and any systematic numerical diffusion will eventually cause the solution to evolve away from the physical solution. A shock capturing scheme like a Roe solver, that comes without any artificial viscosity terms, has minimal diffusion across large pressure jumps.

## 2.1 Initial conditions

The initial conditions for the simulations were described in detail in Paper IV, and therefore we give here only a short summary. The computational grid is in spherical polar coordinates, and runs radially outward from $r_0$ with grid cells of size $\Delta r$ and runs tangentially from the pole ($\theta = 0$) to the equator ($\theta = \pi/2$). Assuming perfect cylindrical symmetry, the full three-dimensional situation can be derived from this grid.

The grid is initially filled with a red giant wind whose density falls off radially as $r^{-2}$ and decreases tangentially from the equator to the pole. This tangential density variation is described by two numbers, $\alpha$ and $\beta$. $\alpha$ determines the density ratio between pole and equator according to

$$\rho_{\text{equator}}/\rho_{\text{pole}} = \frac{1}{1-\alpha} , \quad (1)$$

$\beta$ determines the shape of the tangential variation. When $\beta$ is smaller than 1, the slow wind has the shape of a thin disk, when it is larger than 3, it has a toroidal shape. See for instance Fig. 1 of Mellema et al. (1991) for an illustration of these initial conditions.

The velocity is taken to be constant throughout the slow wind ($v_0$) and the temperature at the polar axis is $T_0$ and follows from pressure equilibrium at other positions. This way the slow wind does not experience any evolution due to the initial conditions.

The two factors that drive the evolution are the fast wind which is kept at a constant mass loss rate ($\dot{M}_{\text{fast}}$) and velocity ($v_{\text{fast}}$) at the inner (radial) edge of the grid, and the stellar radiation output, which is presumed to have a blackbody spectrum (characterized by a luminosity $L$ and an effective temperature $T_{\text{eff}}$) which is also kept constant. In real PNe both the fast wind and the stellar spectrum change in time. This influences the interaction process as can be seen from the results for spherical nebulae in Papers II and III. However, to disentangle the various effects that play a role in shaping PNe, we first consider the simpler case of a non-evolving star and fast wind. The results from Paper IV have shown that these type of models are capable of reproducing observed morphologies and spectra. A future paper will address the effects of these time-dependent boundary conditions (Mellema 1994b, see also Mellema 1993, Ch. 6).

The actual parameters used in the simulations presented in this paper are listed in Table 1. We refer to Paper IV for the exact definitions of the various parameters.

**Table 1.** Input parameters for the simulations.

| Run | A | B |
| --- | --- | --- |
| $\dot{M}_{\text{slow}}$ ( $M_\odot$ yr$^{-1}$) | $5.0\,10^{-6}$ | $5.0\,10^{-6}$ |
| $A$ | 0.9 | 0.7 |
| $B$ | 1.0 | 3.0 |
| $v_0$ (m s$^{-1}$) | $2.0\,10^4$ | $1.0\,10^4$ |
| $T_0$ (K) | $2.0\,10^2$ | $2.0\,10^2$ |
| $\dot{M}_{\text{fast}}$ ( $M_\odot$ yr$^{-1}$) | $5.0\,10^{-8}$ | $1.0\,10^{-7}$ |
| $v_{\text{fast}}$ (m s$^{-1}$) | $2.0\,10^6$ | $2.0\,10^6$ |
| $T_{\text{fast}}$ (K) | $5.0\,10^2$ | $1.0\,10^4$ |
| $T_{\text{eff}}$ (K) | $5.0\,10^4$ | $5.0\,10^4$ |
| $L$ ($L_\odot$) | $8.0\,10^3$ | $1.2\,10^4$ |
| grid dimension | $70 \times 70$ | $80 \times 80$ |
| $r_0$ (m) | $3.0\,10^{14}$ | $1.7\,10^{14}$ |
| $\Delta r$ (m) | $3.0\,10^{13}$ | $2.0\,10^{13}$ |

## 3 COMPARISON WITH NON-RADIATIVE SIMULATIONS

Since non-radiative simulations are relatively easy and certainly much more economical to do, it is interesting to compare radiative and non-radiative simulations. In what ways are non-radiative models good approximations to the more sophisticated radiative models?

Figures 1 and 2 show a time sequences of density plots from a radiative run (A) and non-radiative run (A-NR) with the same initial conditions. The figures show the classic interacting winds situation with an outer shock travelling into the slow wind, the swept-up shell, separated from the low density hot bubble by a contact discontinuity, and (in some of the pictures) the inner shock in the centre.

One can clearly recognize the similarities between the two runs. As in the models shown in Paper I the radiative run evolves somewhat slower because of radiative losses from the hot bubble, but on the basis of these figures one might tend to conclude that there are few differences. This is only partly true, as we will now explain.

### 3.1 Density differences

In the non-radiative simulation the outer shock is a strong shock with the corresponding density jump of 4 ($=(\gamma+1)/(\gamma-1)$). This means that the density in the swept-up shell approximately varies in the same way as the slow wind density, with the highest density at the equator. Since it is in pressure equilibrium with the hot bubble, the shell temperature follows the reverse trend, with the highest temperatures at the pole. This behaviour is modified by the introduction of radiative heating and cooling.

Under equilibrium conditions the balance between heating and cooling tends to keep the gas at a temperature of about 8000 K. This is what happens with the unshocked slow wind material after ionization. The material in the swept-up shell is not necessarily in equilibrium, since it was heated when passing through the outer shock. Whether or not it attains equilibrium depends on the cooling distance, which depends largely on the density.

In the denser equatorial region the cooling distance is very short and in effect unresolved in our simulations, the shock shows up as isothermal. In the less dense polar region, the cooling distance is larger and the swept-up shell has a higher temperature of about 20 000 K.

As said above, the state of the swept-up shell should be such that it is in pressure equilibrium with the hot bubble. Because the radiative heating and cooling play such an important role in determining the temperature of the gas, this requirement has serious consequences for the density distribution in the swept-up shell. In all parts of the swept-up shell that have reached thermal equilibrium, the density should be approximately the same. This is indeed what is seen in run A. Even in the polar areas, where the temperature is slightly above the equilibrium value, the difference is small enough to make the densities similar to the equatorial values. This all means that the actual nebula is a much more homogeneous structure than seen in the non-radiative case. For example, in the run A the lower density polar areas are only 4 times less dense than the equatorial areas (in contrast with the non-radiative simulations where it is a factor of 10), making the polar areas much better observable than would be expected from the non-radiative runs.

An investigation of the shock strengths along the outer shock in the radiative run shows that near the equator the shock is quite weak, with a compression ratio of only 2, whereas at the poles the shock is strong with a compression ratio of about 10 (see Fig. 3, where crosscuts of the density along the pole and equator for both the radiative and non-radiative runs are plotted). The reason for this change is the temperature and hence pressure increase of the slow wind due to the heating by photo-ionization. For the shock to be strong the pressure in the hot bubble should be much higher than in the slow wind. In the radiative case this is no longer true, at the equator the hot bubble pressure and the ionized slow wind pressure are comparable. At the pole, the slow wind density and pressure are 10 times lower, hence the strong shock at that position.

Another way to look at the shock strengths is actually to consider the fact that the pressure in the swept-up shell ($p_{\rm sh}$) is approximately equal to the hot bubble pressure ($p_{\rm hb}$), which is determined by the fast wind properties: $p_{\rm hb} \approx \frac{1}{2}\rho_{\rm fw} v_{\rm fw}^2$. At the same time the temperature in the swept-up shell is determined by cooling and for the high densities at the equator the shock is isothermal at about $T_{\rm sh} = 8500$ K. Because $p \propto \rho T$, $T_{\rm sh}$ and $p_{\rm hb}$ completely determine the shell density $\rho_{\rm sh}$. This means that if the outer shock had been fully isothermal at all positions, $\rho_{\rm sh}$ would have been constant. In that case the compression ratio across the shock ($R = \rho_{\rm sh}/\rho_{\rm sw}$) only varies angularly because the slow wind density does ($\rho_{\rm sw} = \rho_{\rm sw}(\theta)$). In other words, $R(\theta)$ is to first order given by

$$R(\theta) = \frac{\rho_{\rm sh}}{\rho_{\rm sw}(\theta)} \approx \frac{\frac{1}{2}\rho_{\rm fw} v_{\rm fw}^2 m_{\rm p}/(kT_{\rm sh})}{\rho_{\rm sw}(\theta)}. \qquad (2)$$

This means that for given $\rho_{\rm fw}$ and $v_{\rm fw}$, $R$ can acquire different values, depending on $\rho_{\rm sw}(\theta)$. Note that the shock still has to satisfy the (isothermal) shock conditions, so the above explanation is only valid as long as velocities can be found to fulfill the conditions of mass- and momentum conservation across the shock.

At the equator the ratio of slow wind velocity over shock velocity is 0.68 and the slow wind Mach number is 1.4, for which case Chevalier & Imamura (1983) found a compression ratio of about 2, in good agreement with our numerical results (see their Table 5 for isothermal shocks).

The low equatorial compression ratio means that at the equator the swept-up shell is not more compact than in the adiabatic case, as can be seen from Figs. 1 and 2. This is potentially important for processes associated with the contact discontinuity, such as the jet formation mechanism, which depend on the structure of the swept-up shell.

### 3.2 Morphology differences

From a morphological point of view the radiative and non-radiative runs are very similar. However, there are some minor differences. At $t = 634$ years the radiative run already shows signs of a concave contact discontinuity at the equator, whereas in the non-radiative run it still has a convex shape. This is due to the relative weakening of the outer shock at the equator in the radiative case described above. One could say that this weakening increases the effective pole to equator contrast in the slow wind. Note also that the distribution of the observed emission may be different because of the more homogeneous density distribution in the swept-up shell noted above.

Another difference is the absence of axial density condensations in this radiative run. Although the contact discontinuity is somewhat modified later on ($t = 792$ years), the spike-like condensation with vortices seen in the non-radiative run (particularly at $t = 317$ years), is completely absent. Axial condensations are seen in some radiative runs, but in a much milder form than in the corresponding non-radiative runs. These condensations are made out of low density, high temperature gas from the swept-up shell. It is unclear why these condensations are less prominent in the radiative case.

The conclusion is that when one is just interested in the shapes of the bubbles, the non-radiative simulations are useful approximations to the radiative ones. Only minor differences in the basic shapes show up. However, when considering the internal structure of the nebula and its observational appearance, non-radiative runs are not useful approximations, since the density distribution in radiative simulations is found to be very different and the temperatures do not simply follow from thermal equilibrium.

## 4 EVOLUTION OF THE SLOW WIND

As has already been explained in the previous section, photo-ionization and the balance between heating and cooling leave the slow wind at a temperature of about 8000 K. This number depends only weakly on the density. The slow wind becomes isothermal but, because of the density variation, not isobaric: a pressure gradient develops between the high and low density parts. Such a pressure gradient smooths out with a time scale determined by the velocity of sound ($v_{\rm s}$). At a distance $r$ this time is $t_{\rm s} \sim \frac{1}{2}\pi r/v_{\rm s}$. For $r = 10^{15}$ m and $T = 10^4$ K ($v_{\rm s} \sim 10$ km s$^{-1}$), $t_{\rm s} \sim 3000$ years. When the bubble takes longer than this to reach this radius, the effective density contrast will be much lower than the initial value. Consequently the asphericity of the bubble will increase at a lesser rate than expected from the initial conditions.

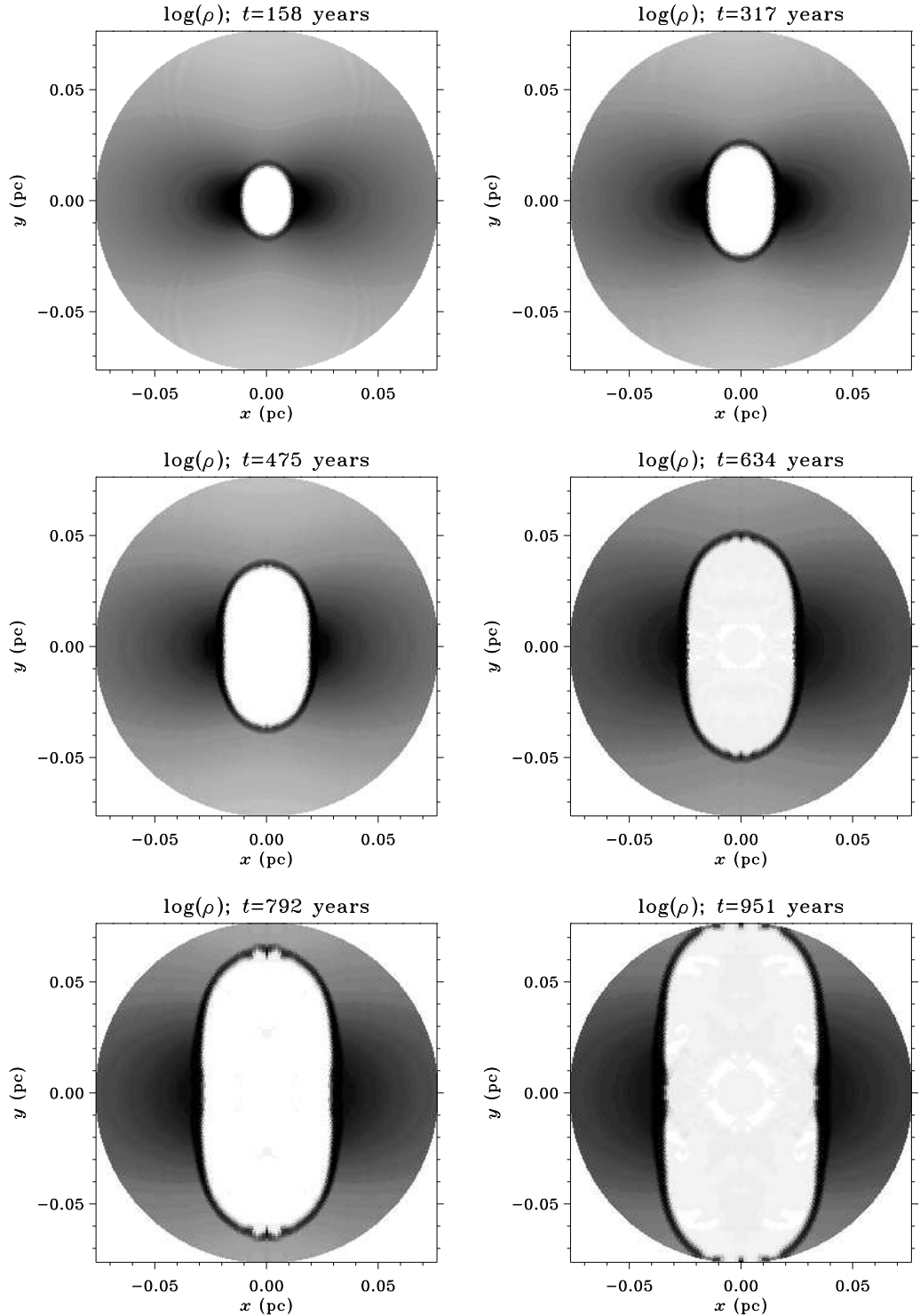

**Figure 1.** Time sequence showing the logarithm of the gas density for run A. Note how the morphology slowly changes in time because the shell expands slower in the equatorial region. The extrema are: $0.93 < n(t = 158 \text{ yrs}) < 3.13\,10^4$ cm$^{-3}$, $0.93 < n(t = 317 \text{ yrs}) < 1.91\,10^4$ cm$^{-3}$, $0.93 < n(t = 475 \text{ yrs}) < 1.61\,10^4$ cm$^{-3}$, $0.098 < n(t = 634 \text{ yrs}) < 1.33\,10^4$ cm$^{-3}$, $0.267 < n(t = 792 \text{ yrs}) < 1.06\,10^4$ cm$^{-3}$, $0.0537 < n(t = 951 \text{ yrs}) < 8.67\,10^3$ cm$^{-3}$.

Figure 4 contains plots from run B which show the density along a curve of constant radius at $t = 0$ and after 634 years of evolution (left) and of the ratio of these two densities (right). In 634 years the effective density contrast has been reduced from 3.3 ($A = 0.7$) to a little over 2.2. The sound speed in the slow wind is 13 km s$^{-1}$, which at $r = 10^{15}$ m results in a sound crossing time from pole to equator of 4000 years. After 15% of this time a reduction of the initial density contrast is already seen. Notice that the density near the equator has increased slightly. This is due to radial movements in the slow wind.

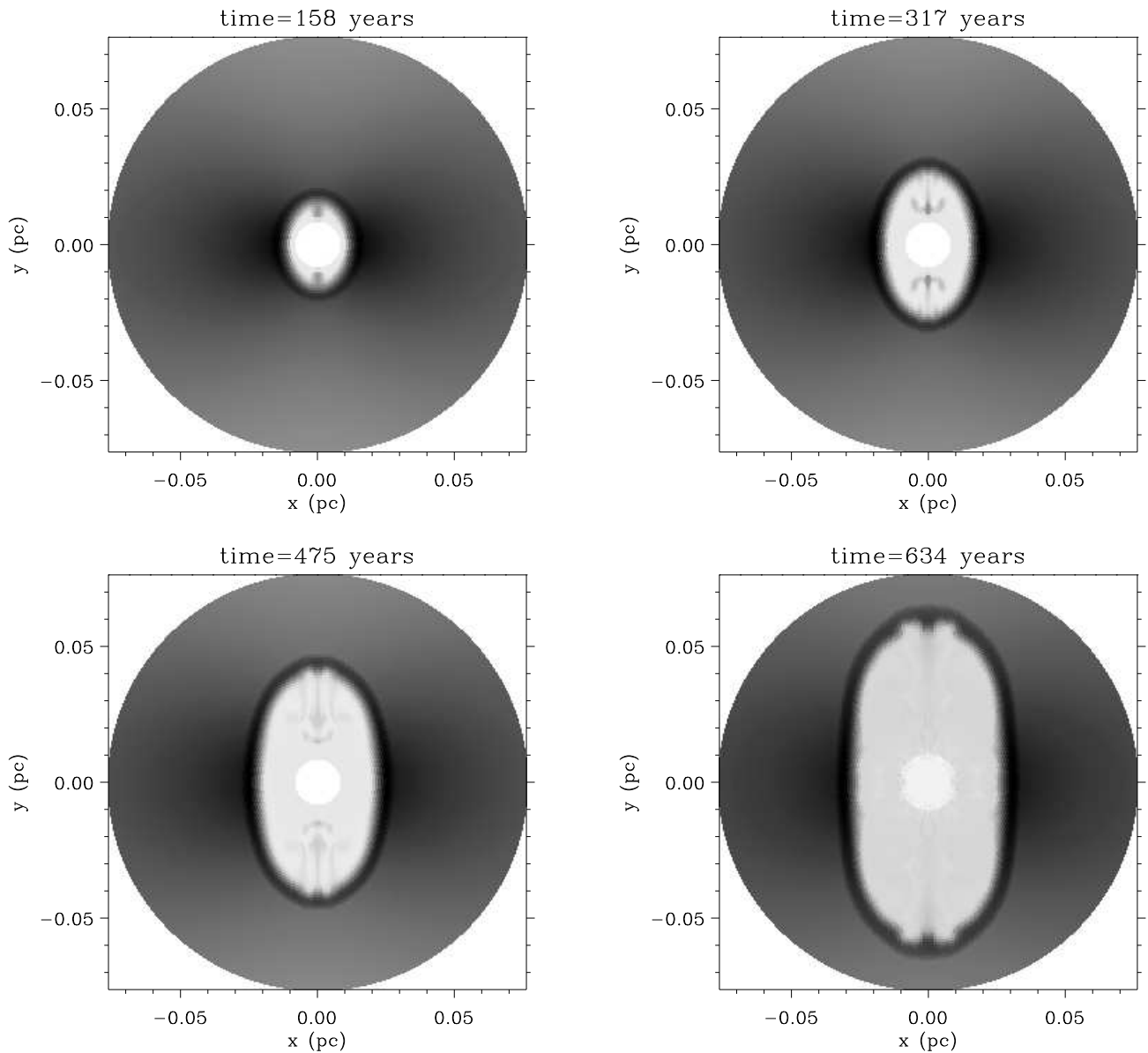

**Figure 2.** Time sequence showing the logarithm of the gas density for run A-NR Compare to run A in Fig. 1. The extrema are: $0.93 < n(t = 158 \text{ yrs}) < 4.50\,10^4$ cm$^{-3}$, $0.93 < n(t = 317 \text{ yrs}) < 3.10\,10^4$ cm$^{-3}$, $0.93 < n(t = 475 \text{ yrs}) < 2.88\,10^4$ cm$^{-3}$, $0.36 < n(t = 634 \text{ yrs}) < 2.38\,10^4$ cm$^{-3}$.

The reason that this behaviour was not seen in the non-radiative simulations is that there the slow wind was explicitly set to be in pressure equilibrium in the tangential direction. This was done to block any evolution of the flow apart from the wind-wind interaction in order to study the aspherical interacting winds model in its simplest form. Because of the inclusion of realistic heating and cooling this is no longer possible.

As a consequence of this slow wind evolution, there is only a limited time to form an extremely aspherical nebula. Within a couple of thousand years after ionization, any original density contrast will have been smoothed out. Any aspherical shell swept up before that time will of course persist. The consequences for the nebula formation are that slowly evolving nebulae are expected to be rounder. A slow nebular evolution is expected in low mass stars, and for fast winds with a low ram pressure ($\dot{M}_{\text{fast}} < 10^{-9}$ M$_\odot$ yr$^{-1}$ for a velocity of 2000 km s$^{-1}$).

## 5 THE STATE OF THE HOT BUBBLE

As can be seen in Fig. 1 the flow inside of the bubble is disordered. This is found to be the case in most simulations.

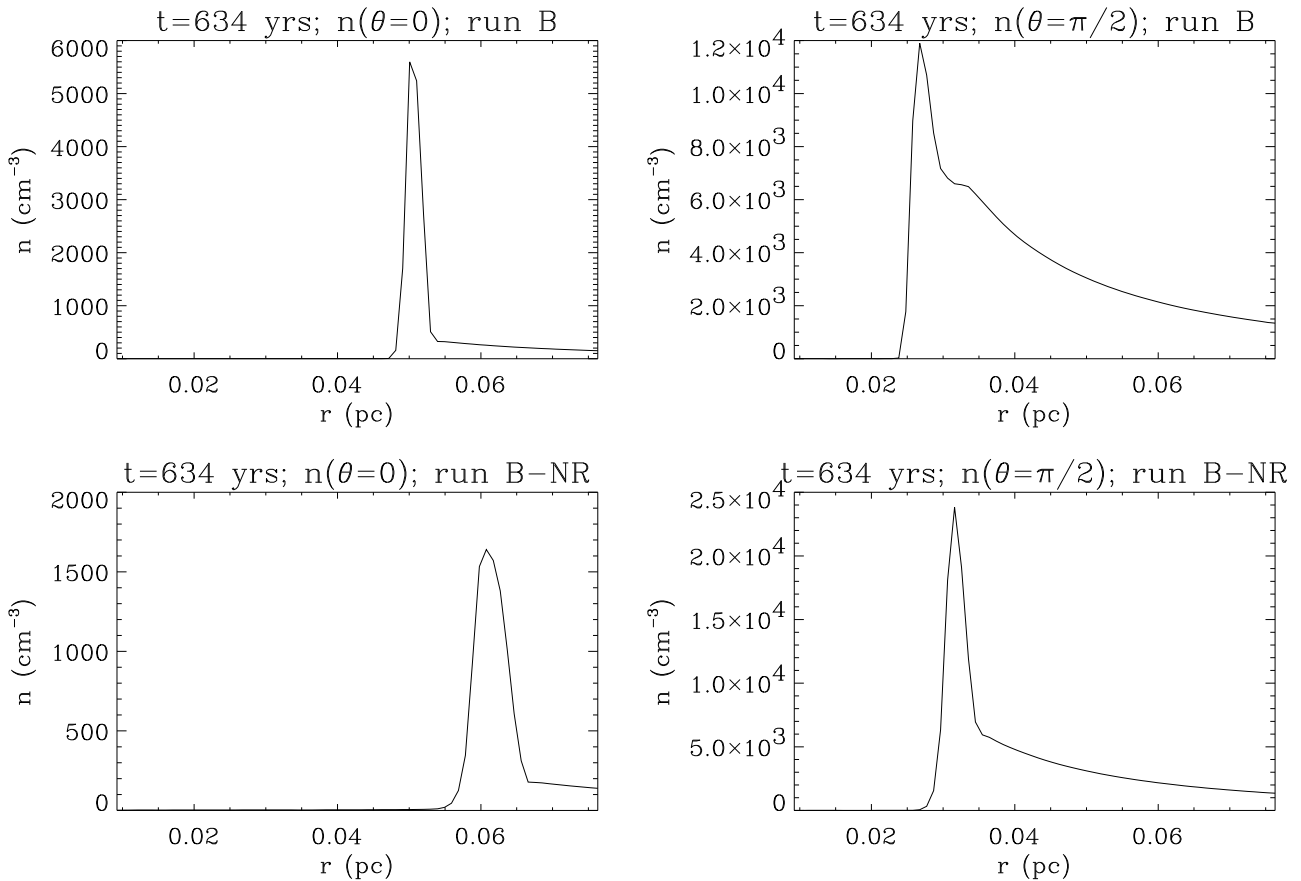

**Figure 3.** Crosscuts of the density distribution along the polar axis ($\theta = 0$) and equator ($\theta = \pi/2$) for run A with (top) and without (bottom) radiative effects. In the radiative case the compression factor along the polar axis is larger than the non-radiative value of 4, whereas along the equator it is less. Note that the compression factor also determines the width of the swept-up shell.

Already in the one-dimensional case waves were found to be oscillating between the inner shock and the contact discontinuity (see Paper I, Fig. 3 and Paper III, Fig. 4). In two dimensions the pattern gets much more complex. The turbulence is triggered by a small corrugation of the inner shock. This corrugation may be related to the 'Odd Even decoupling' a numerical effect reported by Quirk (1992), see Appendix A. Because of this corrugation the fast (radial) outflow goes through an oblique shock at some positions along the inner shock. Only the component of the velocity perpendicular to the shock is discontinuous, while the component parallel to the shock remains unaffected, leading to the formation of high velocity disturbances in the hot bubble. Figure 5 illustrates the formation of these structures. The waves generated in this way travel quickly through the hot bubble and interact with each other to make the bubble turbulent. Because the waves are strongly reflected by the the inner shock and the contact discontinuity they oscillate back and forth between them. Because they are continuously being generated at the inner shock the bubble becomes turbulent and vortices develop. This in turn affects the shape of the inner shock itself.

It is important to note that this behaviour is not an instability. Finger-like structures as shown in Fig. 4 often indicate a Rayleigh-Taylor instability in which, under the influence of an effective gravity, layers of heavier and lighter gas start penetrating each other. But here the high velocity fingers are composed of hot, shocked fast wind material and are not unshocked fast wind material penetrating the hot bubble. Also, although it is mildly corrugated and deformed by the turbulence, the inner shock does not break up as one would expect for a true instability. So, there is no instability, only turbulence caused by waves generated at the inner shock and travelling back and forth in the hot bubble. This situation is reminiscent of the spontaneous emission of sound by a discontinuity described by Landau & Lifshitz (1987), p. 338. Ripples on the surface of the discontinuity (here the inner shock) continue to emit waves without being either damped or amplified.

The turbulence is seen not only in radiative runs, but also in the non-radiative case. This should be so since it is not caused by radiative effects. The non-radiative simulations in Mellema et al. (1991) and Icke et al. (1992) did not show it because the 'fast wind' used there is actually a 'hot wind'. That is, the inner boundary condition used in those simulations is a shocked fast wind (see Mellema et al. 1991, Sect. 3.1). This means that high velocity fingers initially cannot form. In the later stages of these simulations the hot wind moves onto the grid, cools adiabatically and accelerates. Then a shock does develop and with it the high velocity fingers. Under radiative conditions the shocked fast wind can cool and a 'hot wind' inner boundary condition is clearly wrong. That is why we used an unshocked fast wind as the inner boundary condition for the simulations in this paper. As a consequence they suffer from the turbulence generated by it.

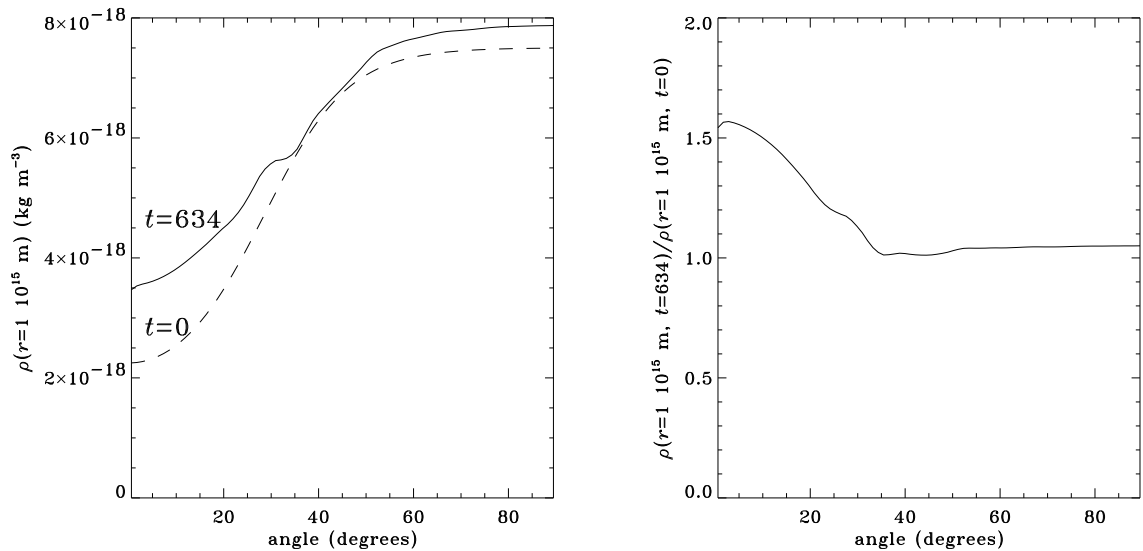

**Figure 4.** Reduction of the initial slow wind density contrast. Results from run B. Left: the density along a curve of constant radius ($r = 1\,10^{15}$ m) for $t = 634$ years (solid line) and $t = 0$ years (dashed line). Right: the ratio of the two curves from the left figure. Notice how the density at the symmetry axis has increased relative to the density at the equator. The effective density contrast has dropped from 3.3 to 2.2.

The velocity suffers most from the turbulence. The pressure does not show much disruption. It does vary, but there are no systematic effects in the variation. Because of the high velocity of sound in the hot bubble, systematic pressure differences are quickly smoothed out. Since it is mainly the pressure of the hot bubble that drives the swept-up shell, the shell is not influenced by the turbulence.

The generation of noise at the inner shock is clearly a numerical artifact. However, the hot bubble's response to it is physical. The high sound speed allows fast travelling waves, which are caught between the inner shock and the contact discontinuity. In real PNe some amount of variation in the fast wind can be expected and therefore the hot bubble will probably become turbulent. Because of the complicated nature of the problem, in which the two solutions for the inner and outer shock have to be matched to obtain the bubble structure, it is not feasible to do an instability analysis. However, the self-similar solutions for the bubble structure presented by Chevalier & Imamura (1983), do show that for the mass loss rates and velocities appropriate for the PN case ($\dot{M}_{\rm slow}/\dot{M}_{\rm fast} \sim 10^{-4}$, $v_{\rm fast}/v_{\rm slow} \sim 10^2$) the inner shock solution is a very sensitive function of these parameters. Figure 5 and Table 4 from Chevalier & Imamura show that for these values, small variations in velocity and mass loss rates lead to large variations in shock position and pressure, indicating that the hot bubble is sensitive to perturbations.

Because the hot bubble produces very little observable emission (see also Sect. 6), it will be hard to prove observationally whether the flow is indeed as turbulent as in the simulations. If the cometary globules seen in the Helix nebula are indeed contained in the hot bubble, this would indicate that in this case the flow is predominantly radial. However the central star of the Helix nebula is observed to no longer have a fast wind and the situation for the globules is not completely clear, so this can not really be used to rule out turbulence (Meaburn et al. 1992; Hartquist & Dyson 1993).

It is clear that the two-dimensional situation offers many more possibilities for turbulence to develop than the one-dimensional case. However, the cylindrical symmetry imposed in the two-dimensional case still limits the development. Because the equatorial plane and the symmetry axis are treated as perfect mirrors, waves reflect against them. This influences the flow, and in particular the inner shock. The radial position of the inner shock near the equator is invariably pinched, hence the 'boxy' appearance of the inner shock in Figs. 1 and 4. The depression of the inner shock position near the equator is caused by a vortex which preferentially forms there and often reaches as far out as the contact discontinuity. This deformation of the inner shock is clearly caused by the imposed numerical boundary condition and can therefore not be trusted. A possible solution for this problem is to extend the grid so that it covers the entire area on one side of the symmetry axis ($0 < \theta < \pi$), thus enabling the flow to cross the equator. Luckily, the effect on the astrophysically interesting parts of the bubble is small: the structure of the swept-up shell is not affected by the turbulence. As was said above, the hot bubble pressure which drives the swept-up shell is never systematically modified.

As a side-effect the turbulence triggers a numerical thermal instability at the inside of the contact discontinuity. Small areas of low density and high temperature (pressure equilibrium is maintained) develop. Because the allowed in-tegration time step is a function of the sound speed, these high temperature spots result in small time steps. In higher resolution simulations (100 by 100 grid points) this slows down the program too much to obtain results within a reasonable amount of computer time. This is the reason why all simulations shown in this paper are done on a rather coarse grid (70 by 70 or 80 by 80 grid points). Here these points also occur but generally do not bring down the time step to unreasonable values. Efforts are being undertaken to prevent this behaviour.

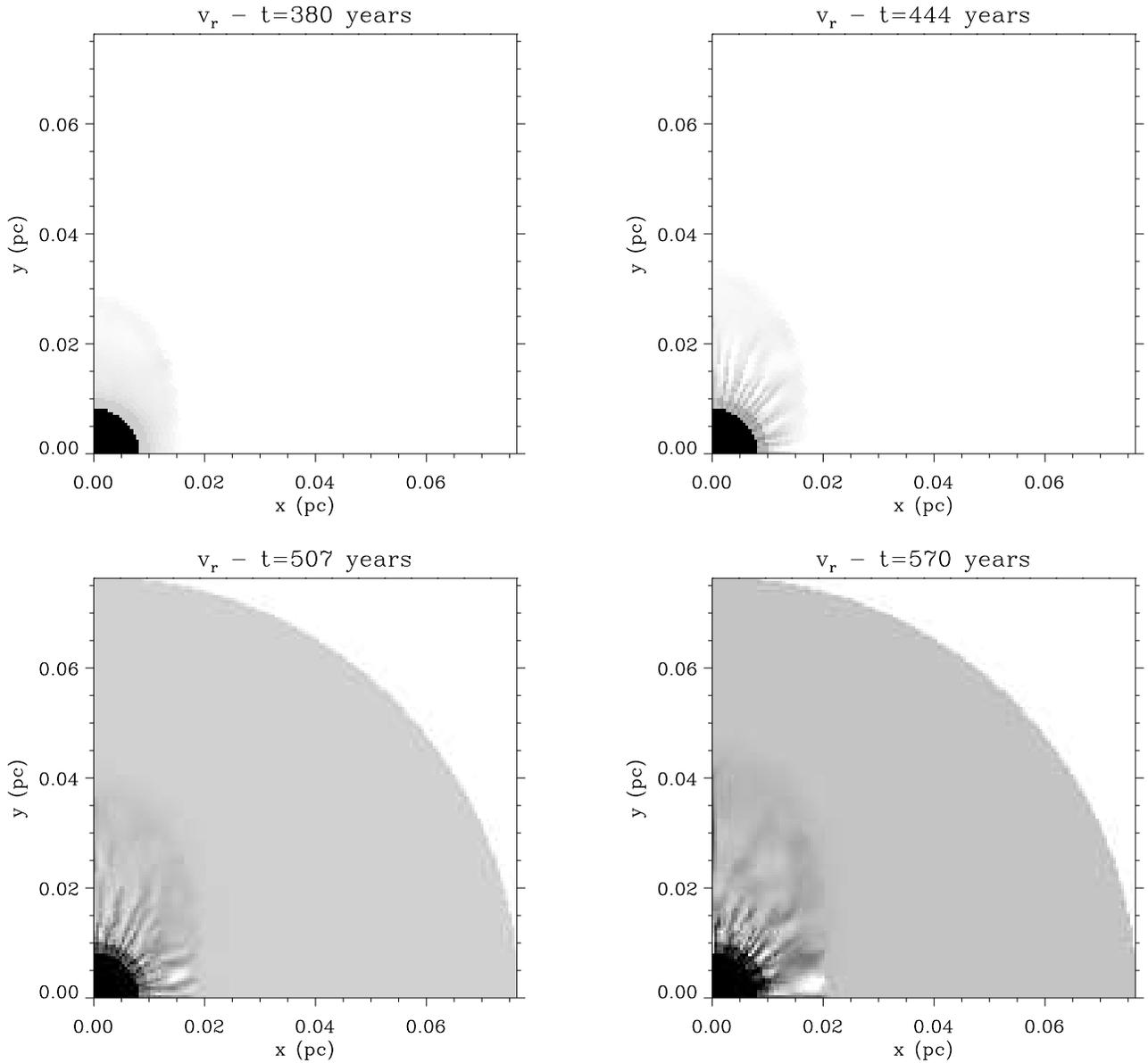

**Figure 5.** Development of the turbulence at the inner shock (run A). At $t = 380$ years the turbulence has not yet developed. At $t = 444$ years high velocity fingers form in the hot bubble, perturbing it at later times. The extrema are: $20 < v(t = 158 \text{ yrs}) < 2.0\,10^3$ km s$^{-1}$, $20 < v(t = 317 \text{ yrs}) < 2.0\,10^3$ km s$^{-1}$, $2.4 < v(t = 475 \text{ yrs}) < 2.0\,10^3$ km s$^{-1}$, $8.3 < v(t = 634 \text{ yrs}) < 2.0\,10^3$ km s$^{-1}$.

## 6  SOFT X-RAY EMISSION

In Paper IV we presented observables in the form of synthesized images and long slit spectra in the lines of important ionized species, such as [OIII]. However, modern observations of PNe span the entire electromagnetic spectrum. Radio observations at 6 cm probe the same ionized material as the hydrogen recombination lines and therefore do not provide us with new information about the PN morphologies and structure. Molecular lines such as the CO rotational transitions and the IR H$_2$ lines probe the neutral and molecular material in the PN system (Huggins 1993). In our models not much can be said about this, since the stellar properties were chosen to ionize all the gas on the grid and in this case molecular material can only survive in dense clumps, which we do not model. These clumps could be the subject of any further studies of the GISW model, but note that simple numerical modelling will not do, since the typical clump size is smaller than a grid cell. There are some observational indications in the form of relatively high velocity molecular outflows, that seem to point to a significant dynamical change to the molecular material after the AGB (Sahai et al. 1994).

ROSAT has supplied us with soft X-ray images of PNe (Kreysing et al. 1992). This emission is expected to probe very different regions of the PN system than the optical lines. Soft X-rays are mainly produced by bremsstrahlung in high temperature regions, such as the hot bubble. Images of bremsstrahlung emission can be constructed from our models and in this section we compare the soft X-rays from our simulations with the actual ROSAT observations.

Figure 6 shows a soft X-ray image from run A at a time $t = 792$ years. One can clearly recognise the shape of the nebula, such as shown in Fig. 1. We approximate the soft X-ray emission by the calculating the bremsstrahlung in the energy range 0.1 to 2.4 kev, the energy range of the ROSAT PSPC instrument. This radiation mainly originates from the interface between the hot bubble and the swept up shell. Here both the density and temperature are high enough to produce appreciable amounts of soft X-ray photons. The coincidence of both these requirements is rare enough to result in a patchy distribution of emission. This is a resolution effect, across the contact discontinuity the temperature changes from $10^4$ to $10^7$ K and the density from $10^4$ to 10 cm$^{-3}$. There are typically some 4 or 5 grid cells across this discontinuity and only those with temperatures of $\sim 10^6$ K and densities of $\sim 10^2$ cm$^{-3}$ produce substantial amounts of soft X-rays. To hide this resolution effect, the image has been smoothed with a Gaussian of FWHM of 1/30 of the size of the image.

The total soft X-ray luminosity coming from the nebula is found to be $1.5\,10^{30}$ erg/s, which amounts to 0.002% of the mechanical luminosity of the fast wind ($\frac{1}{2}\dot{M}v^2 = 6.3\,10^{34}$ erg/s $= 16\mathrm{L}_\odot$). Because of line emission the actual flux may be somewhat higher.

Other runs produce similar types of results, leading to the conclusion that the soft X-ray emission from PNe is expected to be coming from a relatively thin shell, just inside of the optical image. The ROSAT PSPC observations presented by Kreysing et al. (1992) can not really be used to say something about the distribution of the soft X-ray emission. In only two cases is the result based on more than a few dozen photons and one of these (NGC 6853) has been reported to be a ghost image (Chu et al. 1993). The other one (NGC 6543) has a size close to PSPC point spread function, which is 30" and increases considerably at off axis angles (Wrigge, private communication). The observations can be used to say something about the temperature of the main source of the photons and for three of the six objects (NGC 6543, BD+30 3639, A 12) this is derived to be around $10^6$ K, in line with our model. The other three (NGC 4361, NGC 6853, LoTr 5) give temperatures of around $10^5$ K, consistent with the main source being the hot central star. For those nebulae in which the central star is known to have a fast wind, the estimates for the soft X-ray luminosity are about 0.1% of the mechanical luminosity of the fast wind. Although this is a somewhat higher efficiency than in our model, it is substantially less than 1, showing that only a small fraction of the fast wind energy input is converted into soft X-rays.

A very similar case to PNe are the wind blown bubbles around Wolf-Rayet stars, such as NGC 6888, the ROSAT observations of which are presented by Wrigge et al. (1994). These authors find that the X-ray emission only originates from a thin filamentary shell located on the inside of the optical H$\alpha$ picture and which only fills of order 1% of the bubble volume. This matches the results from our model. The total soft X-ray luminosity is about 0.03% of the mechanical wind luminosity, somewhat more than in the model, but again substantially less than the fast wind mechanical luminosity.

Wrigge et al. compare their results to the (analytical) models of Weaver et al. (1977) and find a discrepancy. The Weaver et al. models predict a filled bubble of X-ray emission, instead the observations show the emission to come from a thin filamentary shell.

The reason our models differ from those of Weaver et al. is that the latter take into account a large amount of thermal conductivity. Because of this, part of the swept up shell evaporates into the hot bubble, thereby increasing its density and lowering its temperature. The result is that the entire hot bubble becomes an efficient soft X-ray emitter. In our numerical models their is no explicit thermal conduction. However, numerical diffusivity across the contact discontinuity has the same effect as a very low amount of thermal conductivity and creates a thin interface of $\sim 10^6$ K which produces the soft X-rays. A small randomly oriented magnetic field suffices to drastically reduce the thermal conductivity (see e.g. Soker 1994). This makes it likely that the thermal conductivity is indeed almost negligible.

Two other solutions can be offered. Wrigge et al. propose that a clumpy distribution of the surrounding medium and the swept-up material might explain the observed difference between observations and theory. The H$\alpha$ images of NGC 6888 indeed point to a clumpy distribution of matter. How a clumpy distribution may lead to the observed characteristics of soft X-ray emission is not discussed. Presumably the soft X-rays mainly originate from the interfaces of hot and cold material, but in this case it is not clear why the emission is distributed the way it is.

Another possibility is that there is only a tiny hot bubble because the flow is momentum conserving rather than energy conserving. There is marginal indication for this in the estimates for values of the energy and momentum of the fast wind and the swept up shell (Wrigge et al. 1994). In the momentum conserving case the fast wind is cooling so efficiently that a hot bubble almost does not form, because most of the energy is lost in the form of radiation. For the cooling to be so efficient the densities must be extremely high, since the kinetic energy content of the fast wind is so large ($v_{\mathrm{fast}} = 2400$ km s$^{-1}$). Based on the estimated mass loss rate and the size of the nebula, the densities can be shown to be too low for such efficient cooling to take place. This effectively rules out the momentum conserving case.

So, the evidence seems to point to low thermal conductivity leading to high temperatures and low densities for the hot bubble of NGC 6888. In absence of similar quality data for PNe this conclusion can only be tentatively be carried

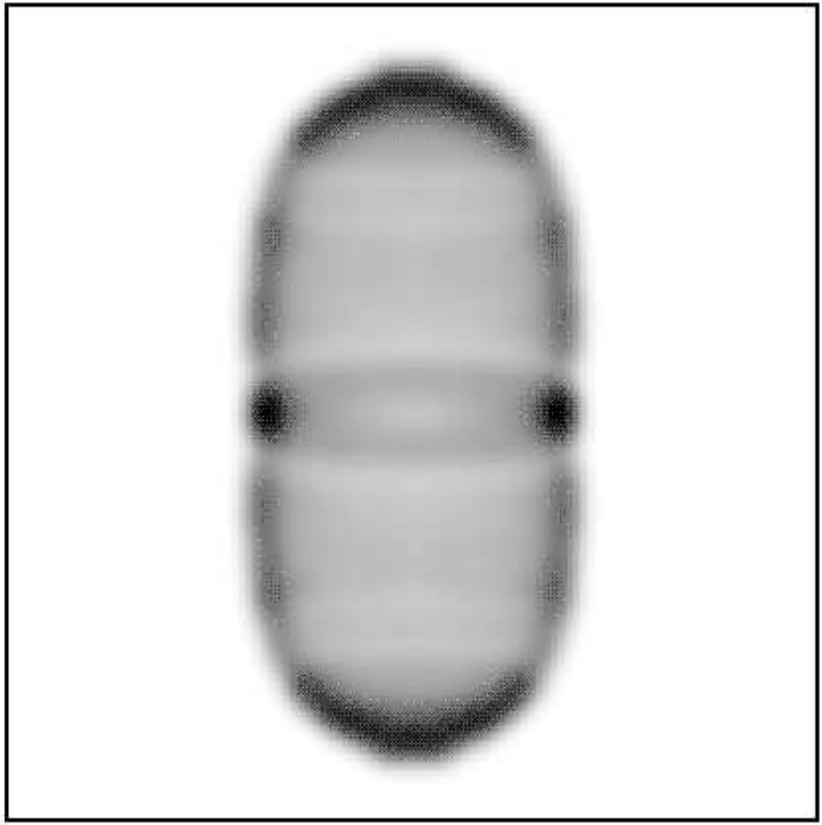

**Figure 6.** Soft X-ray emission from run A at $t = 792$ years. The image has been smoothed with a Gaussian of FWHM of 1/30 of the size of the image.

over to them.

## 7  CONCLUSIONS

This paper deals with various aspects of the Generalized Interacting Winds Model for aspherical PNe. The unifying theme is the physical structure of the nebula and its surroundings, something which was not addressed in the previous article, which concentrated more on the morphologies. The numerical models show that

1) The radiative effects of heating and cooling do change the shock structure and internal density distribution of the nebulae. After the surrounding slow wind has been ionized, the outer shock near the equator is likely to be weak.
2) In areas were the density is high enough for cooling to be efficient, the nebula is expected to be of roughly constant density. This in turn means that these parts of the nebula are equally bright, which is very different from what is found in the non-radiative case.
3) In low density regions with inefficient cooling, the temperature is higher due to shock heating. This means that there is no unique nebular temperature, just as there is no unique nebular density. This may have consequences for standard nebular analysis based on line ratios and explain why different methods give different values for the electron temperatures (Liu & Danziger 1993).
4) After ionization the asphericity in the slow wind is slowly eroded away because of pressure waves. This puts an upper limit on the time scale for the formation of aspherical PNe: if the PN takes longer than 3000–4000 years to form, it will not be very aspherical.
5) The hot bubble is very sensitive to small disturbances and becomes turbulent in most simulations. Although the triggering of the turbulence is a numerical effect, variations in the fast wind properties are expected to turn the hot bubble turbulent in real PNe. This will be hard to observe.
6) Due to a low thermal conductivity, soft X-rays are only expected from a thin interface between the nebula and the hot bubble. ROSAT observations of NGC 6543 and the Wolf-Rayet nebula NGC 6888 appear to confirm this. This means that wind-blown bubbles are inefficient soft X-ray producers, with less than 0.1% of the fast wind energy coming out in the form of soft X-rays.

## ACKNOWLEDGEMENT

The authors would like to thank Matthias Wrigge for his help with interpreting the ROSAT results on PNe and appreciate the suggestions of Robin Williams and Alex Raga for improving the paper. GM is supported by a PPARC research assistantship.

## APPENDIX: ODD-EVEN DECOUPLING

Quirk (1992) has shown that the Roe solver together with several other Riemann solvers suffers from what he calls 'Odd Even decoupling'. In high resolution simulations of a slow moving shock, Quirk found that it at some point starts to show spike-like disturbances, in which the pressure disturbances are out of phase with the density disturbances. This behaviour is also seen in the 'fingers' discussed in Sect. 5. However, these are not high resolution simulations, and therefore it is not entirely certain that this is the same behaviour.

Simulations using the LCD/FCT method (Boris & Book 1973; Icke 1991) show similar disturbances being generated at the inner shock. However, the more diffusive character of this method ensures that these disturbances disappear in time either at the contact discontinuity, or by interacting with themselves.